\documentclass{emulateapj}

\usepackage{graphicx}
\usepackage{amsmath}
\usepackage{ulem}
\usepackage{color}

\newcommand{\feoh}{[{\rm Fe} / {\rm H}]}
\newcommand{\msun}{\, M_\odot}
\newcommand{\Zsun}{\, Z_\odot}
\newcommand{\mmd}{m_{\rm md}}

\newcommand{\Tv}{T_{\rm vir}}

\newcommand{\christlieb}{HE 0107-5240}
\newcommand{\keller}{SMSS 0313-6708}

\newcommand{\frebel}{HE 1327-2326}

\def\abra#1#2{[{\rm #1}/{\rm #2}]}

\newcommand{\mhyph}{\, \mathchar`- \, }

\shorttitle{Pop~III stars}
\shortauthors{Komiya et al.}

\begin{document}
\title{The most iron-deficient stars as the polluted population III stars}

\author{Yutaka Komiya\altaffilmark{1}, Takuma Suda\altaffilmark{1} and Masayuki Y. Fujimoto\altaffilmark{2}\altaffilmark{3}}
\altaffiltext{1}{Research Center for the Early Universe, University of Tokyo, Bunkyo-ku, Tokyo, Japan}
\altaffiltext{2}{Department of Cosmoscience, Hokkaido University, Sapporo, Hokkaido 060-0810, Japan}
\altaffiltext{3}{Department of Engineering, Hokkai-Gakuen University, Sapporo, Hokkaido 062-8605, Japan}

\begin{abstract}
We investigate the origin of the most iron-poor stars including SMSS J031300.36-670839.3 with $\feoh < -7.52$.  
   We compute the change of surface metallicity of stars with the accretion of interstellar matter (ISM) after their birth using the chemical evolution model within the framework of the hierarchical galaxy formation. 
   The predicted metallicity distribution function agrees very well with that observed from extremely metal-poor stars.   
   In particular, the lowest metallicity tail is well reproduced by the Population III stars whose surfaces are polluted with metals through ISM accretion. 
   This suggests that the origin of iron group elements is explained by ISM accretion for the stars with $\feoh \lesssim -5$.   
   The present results give new insights into the nature of the most metal-poor stars and the search for Population III stars with pristine abundances.  
\end{abstract}

\section{Introduction}

The most metal-poor stars are relics of stars formed in the early universe. 
If low-mass stars with mass $m \lesssim 0.8 \msun$ were born from the primordial gas, they would still exist as nuclear burning stars in our nearby field. 
It is still controversial whether there are such low-mass, metal-free stars, which we refer to as Population (Pop) III survivors in this paper.  
Recent simulations, however, raise the possibility that low-mass stars were formed in the pristine environment \citep[see review by][]{Bromm13}.  

Observationally, Pop~III survivors have not yet been identified despite longstanding efforts \citep[see][]{Bond81}.  
   Thanks to large-scaled surveys in the Milky Way (MW) halo from the 1990s \citep[][for reviews]{Beers05}, several hundreds of stars with the metallicity of $\feoh < -3$ are known to date and studied with high resolution spectroscopy \citep[see e.g.][]{Suda08, Suda11}. 
   Among them, six stars were found below $\feoh < -4.5$ but with finite metallicity.  
 For the most iron-poor star, SMSS J031300.36-670839.3 \citep[][hereafter \keller]{Keller14}, iron lines are not detected ($\abra{Fe}{H} < -7.52$, Bessell et al. 2015) while calcium abundance is determined to be $\abra{Ca}{H} = -7.26$.   
   Two other stars have $\feoh = -5.7$ \citep[\frebel,][]{Frebel05} and $-5.4$ \citep[\christlieb,][]{Christlieb02}, respectively.  
   The remaining three lie between $-5 < \feoh < -4.5$ \citep{Caffau11,Norris07,Hansen14}. 
   We refer to stars with $\feoh \le -5$ and $-5 < \feoh \le -4.5$ as hyper and ultra metal-poor (HMP and UMP) stars, respectively \citep[c.f.][]{Beers05}. 

In order to explain the absence of Pop~III survivors, surface pollution with accreted metals from the interstellar matter (ISM) has been proposed \citep{Yoshii81, Iben83}. 
   It is also argued for the extremely low metallicity of HMP stars \citep{Shigeyama03,Suda04,Komiya09L}.  
   We investigated the change of surface iron abundance of Pop~III survivors through ISM accretion using chemical evolution modelling within the framework of hierarchical structure formation \citep{Komiya09L,Komiya10}, and demonstrated that Pop~III survivors can be polluted up to $\feoh \sim -5$. 
   The discovery of the star with metallicity smaller by two orders of magnitude gives us a good reason to discuss the range of surface metallicity of polluted Pop~III survivors.  

In this paper, we revisit the problem of surface pollution using an updated chemical evolution model \citep{Komiya14}.  
   Our purpose is to explore the metallicity distribution function (MDF) at extremely low metallicity and discuss the origin of HMP/UMP stars. 
 We focus on the iron abundance since lighter elements such as carbon and magnesium can be influenced by binary mass transfer \citep{Suda04, Komiya07, Nishimura09}.

\section{Computation Method} 

\subsection{The Hierarchical Chemical Evolution Model}

We compute the chemical evolution of the MW within the framework of the hierarchical structure formation in the $\Lambda$ cold dark-matter universe to evaluate the surface pollution of Pop~III survivors.   
   The detailed description of our model is given in \citet{Komiya14}, and we review the characteristics relevant to the surface pollution below.

We build merger trees based on the extended Press-Schechter theory \citep{Lacey93, SK99}. 
   The total mass of the MW is taken to be $2 \times 10^{12} \msun$ \citep{Li08, Boylan-Kolchin13}.  
   The low-mass limit of mini-halos is determined by the virialized temperature $\Tv = 10^3\ {\rm K}$ \citep{Tegmark97, Yoshida03}.

The star formation rate is assumed to be proportional to the gas mass, $M_{\rm gas}$, of mini-halos with the  star formation efficiency (SFE), $\epsilon_{\star}$, set constant.  
   We register all the individual stars with metallicity $Z < 0.1\Zsun$.  
   We adopt a log-normal form of the initial mass function (IMF). 
   The typical mass, $\mmd$, of EMP stars is estimated at $\mmd = 2.5 \mhyph 20\msun$, much higher than the present-day IMF, according to the statistics of carbon-enhanced stars among EMP stars, which arises through the mass transfer from AGB stars in binary systems \citep{Komiya07,Komiya09,Suda13}.  
   We adopt $\mmd = 10 \msun$ for EMP stars and $25 \msun$ for Pop~III stars with the dispersion of $\sigma = 0.4$, and $\epsilon_{\star} = 10^{-11}{\rm yr}^{-1}$ as fiducial values from the previous study \citep[][see Table~\ref{Tparam}]{Komiya10}. 
   One of the consequences of the high-mass IMF is that EMP and Pop~III survivors mostly form as the secondary members in binary systems \citep{Komiya07}. 
   We take into account that the Lyman-Werner (LW) background prohibit star formation in the mini-halos with $\Tv<10^4$ K formed later than $z = z_{\rm LW} = 20$, although those formed earlier continue their star formation \citep{Ricotti02, Ahn07}.

\begin{table*}
\begin{center}
\caption{Parameters and their fiducial value}
\label{Tparam}

\begin{tabular}{|c|l|c|c|}
\hline
parameter  & description & fiducial value  \\
\hline
$\epsilon_{\star}$	& star formation efficiency	& $10^{-11} {\rm yr^{-1}}$	\\
$m_{\rm md, emp}$	& median mass of IMF for EMP stars	& $10\msun$	\\
$m_{\rm md, p3}$	& median mass of IMF for Pop~III stars	& $25\msun$	\\
$\sigma$	& variance of IMF	& 0.4	\\
$\eta$			& kinetic energy fraction for SNe	& 0.1	\\
$z_{\rm LW}$	& redshift of Lyman-Werner radiation	& 20	\\
$f_b$	&		 binary fraction 			& 0.5 \\
$n(q)$			& mass ratio distribution of binaries	& 1	\\
$\dot{m_1} / \dot{m_2}$	& ratio of ISM accretion rate on each binary component	& 1	\\
 \hline
\end{tabular}
\end{center}
\end{table*}

   We assume a uniform mixing of SN ejecta in the mini-halos. 
   The gas outflow from the mini-halos is computed for each SN and HII region as a function of their kinetic energy and the binding energy of the mini-halos. 
The conversion efficiency, $\eta$, of the SN explosion energy to kinetic energy is set at $0.1$. 
   It is considered to form a galactic wind that enriches the intergalactic medium (IGM) with metals. 
   The evolution of the metal-enriched wind is followed by assuming the momentum conserving snow-plow shell model.

Some of the mini-halos are formed with IGM pre-enriched with metal by the galactic wind.  
   In this study, we improve the treatment of the pre-enrichment by considering the distance between the mini-halos, while a random spatial distribution was assumed in the previous work. 
   The $m$-th mini-halo is regarded as pre-enriched by a galactic wind from the $n$-th halo if the distance, $D_{n,m}$, between them is smaller than the radius, ${\cal R}_n$, of the metal enriched region around the $n$-th halo at formation.  
   The distance $D_{n,m}(t)$ is estimated following the formalism of the EPS theory in which the formation of a halo is described as the gravitational collapse of a spherical overdense region. 
   Two halos are considered to ``merge'' when a halo is formed by gravitational collapse with the two halos incorporated. 
   In this framework, the distance between the two halos that merge at time $t_{\rm merge}$ to form one halo of total mass $M_{\rm tot}$ may be approximated to the radius, $r_{}$, of the spherical region of over-density with mass $M_{\rm tot}$ that collapses at $t_{\rm merge}$.

\subsection{Surface Pollution}\label{Spollution}

We trace the change in the surface iron abundance of each star by ISM accretion. 
   In this work, we also take into account the transfer of iron from the surface of a primary to the secondary member in the binary system through the stellar wind.  
   The accreted metals are assumed to be mixed in the surface convection zones with $0.2\msun$ for giants and of $0.0035\msun$ for dwarfs \citep{Fujimoto95}.

For the ISM accretion, we adopt the Bondi-Hoyle accretion formula 
\begin{equation}
\dot{m} = \frac{4\pi (Gm)^2 \rho_{\rm gas}}{(V_r^2+c_s(T_{\rm gas})^2)^{3/2}} 
\label{mdotEq}
\end{equation} 
   with the gas density, $\rho_{\rm gas}$, the sound velocity, $c_s(T_{\rm gas})$, as a function of the gas temperature, and the relative velocity, $V_r$, of the star to the ambient gas, and the stellar mass $m$.

We assume that gas in a mini-halo is concentrated in its central region via isobaric contraction, i.e., 
\begin{equation}\label{eq:rho200}
 \rho_{\rm gas}=\rho_{\rm vir }(z) \left(\frac{M_{\rm gas}}{M_{\rm h}}\right) \left(\frac{T_{\rm gas}}{T_{\rm vir}} \right)^{-1}, 
\end{equation} 
where $\rho_{\rm vir}$ is the averaged density of the virialized dark-matter halos. 
   We set $T_{\rm gas}=200$ K and 50 K for the neutral and ionized primordial clouds, respectively \citep{Uehara00}. 
   When the gas metallicity exceeds $\abra{Z}{H} = -6$, we change the temperature to $T_{\rm gas}={\rm max}(10 {\rm K}, T_{\rm CMB}(z))$, where $T_{\rm CMB}$ is the temperature of the cosmic microwave background, since dust cooling becomes effective \citep{Omukai05, Schneider06}.

We set $V_r = c_s$ before the host halo of the star undergoes a merger event since stars are also thought to be centrally concentrated. 
   After the merger, stars from the smaller halo will be scattered through the merged halo with little dissipation of their kinetic energies, and we set $V_r = V_{\rm circ}$, where $V_{\rm circ}$ is the circular velocity of the merged halo.  
   The density $\rho_{\rm gas}$ is replaced by $\rho_{\rm vir } ({M_{\rm gas}}/{M_{\rm h}})$ taking into account the filling factor of gas clouds in the halo. 

In the case of binary systems, the accretion rate on each member is still poorly understood.  
   For simplicity, we take an equal accretion rate for both components, i.e., $\dot{m_1}/\dot{m_2} = 1$, in the fiducial model. 

The mass transfer rate in the binary systems is computed as a function of stellar mass and binary period with a given wind velocity from the Bondi-Hoyle formulae as in \citet{Komiya09}.

\section{Results}\label{Sresult}

\begin{figure}
\includegraphics[width=\columnwidth]{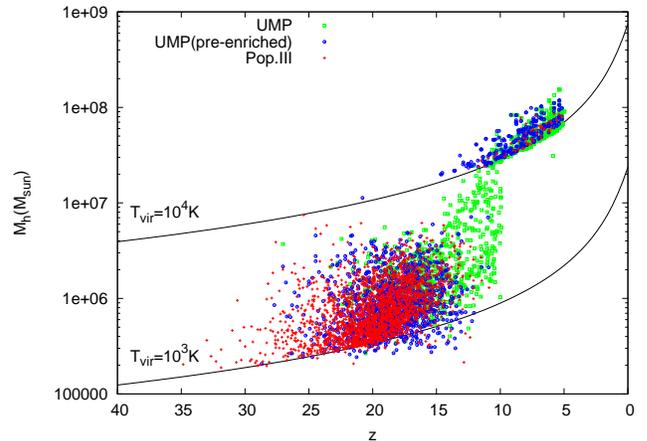}
\caption{
The formation redshift and the mass of host mini-halos for Pop~III survivors (red crosses), the second generation UMP stars (green squares) and the first generation UMP stars in the pre-enriched mini-halos (blue circles).  
Lines denote the mass of mini-halos with the virial temperatures attached.  
}\label{z-m}
\end{figure}

Figure~\ref{z-m} shows the formation redshift and host halo mass of Pop~III and UMP survivors for the model with the fiducial parameters given in Table~\ref{Tparam}. 
In this model, $2100 \mhyph 2500$ low-mass Pop~III stars are formed in mini-halos with $\sim 10^6 \msun$ around $z \simeq 20$.  
Most of the second generation stars have metallicity of $\feoh > -4$ since the iron from one SN of $\sim 0.07\msun$ is mixed with the gas of $\sim 10^5\msun$.  
After $z = z_{\rm LW} = 20$, newly formed mini-halos are prevented from forming stars by the LW radiation, while those formed at earlier epochs, two thirds of which are still metal-free, continue to form stars.  
The mini-halos later formed grow in mass to have $T_{\rm vir} > 10^4\ {\rm K}$ by $z \lesssim 10$ and form $\sim 100$ Pop~III survivors. 
The second generation stars in these massive halos have lower metallicity of $-5 \lesssim \feoh \lesssim -4$ because SN ejecta is mixed homogeneously with a larger mass of gas. 
The first generation of stars in mini-halos with pre-enriched IGM have metallicity around $-6 \lesssim \feoh \lesssim -2$. 
They account for half of the UMP stars.

\begin{figure}
\includegraphics[width=\columnwidth]{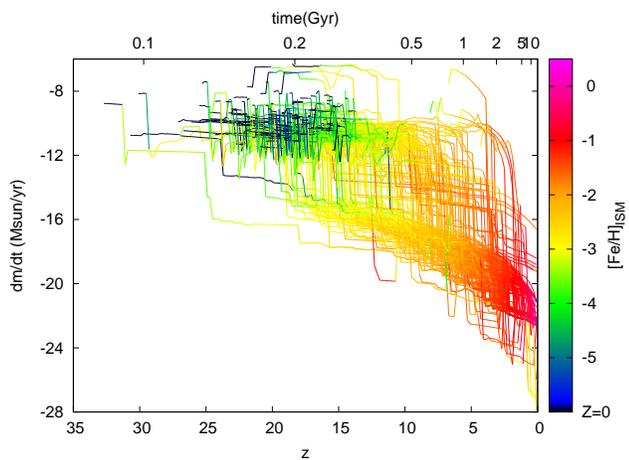}
\caption{
Variations of the ISM accretion rate for the sampled Pop~III survivors plotted against the redshift. 
The metallicity of accreting gas is color-coded. 
}\label{accretion}
\end{figure}

Figure~\ref{accretion} illustrates the ISM accretion rate with the metallicity of accreting gas color-coded as a function of the redshift for the sampled Pop~III survivors.  
The accretion rate is much higher in the host mini-halos than in the merged halos because $V_r$ increases after a merger event.   
Typically Pop~III survivors accrete $\sim 10^{-8.5} \msun$ of iron to have the surface metallicity of $\feoh \sim -5$ for giant stars, and $\feoh \sim -3$ for dwarfs. 
The surface metallicity of Pop~III survivors spans a wide range between $-8 \lesssim \feoh \lesssim -2$ depending on the merging history of host halos and the mass of the primary stars of Pop~III survivors.  
 The abundance pattern of the polluted Pop~III stars should be similar to those of EMP stars because accretion in the early universe with $\feoh \lesssim -3$ is more efficient than metal-rich counterparts. 

\begin{figure}
\includegraphics[width=\columnwidth]{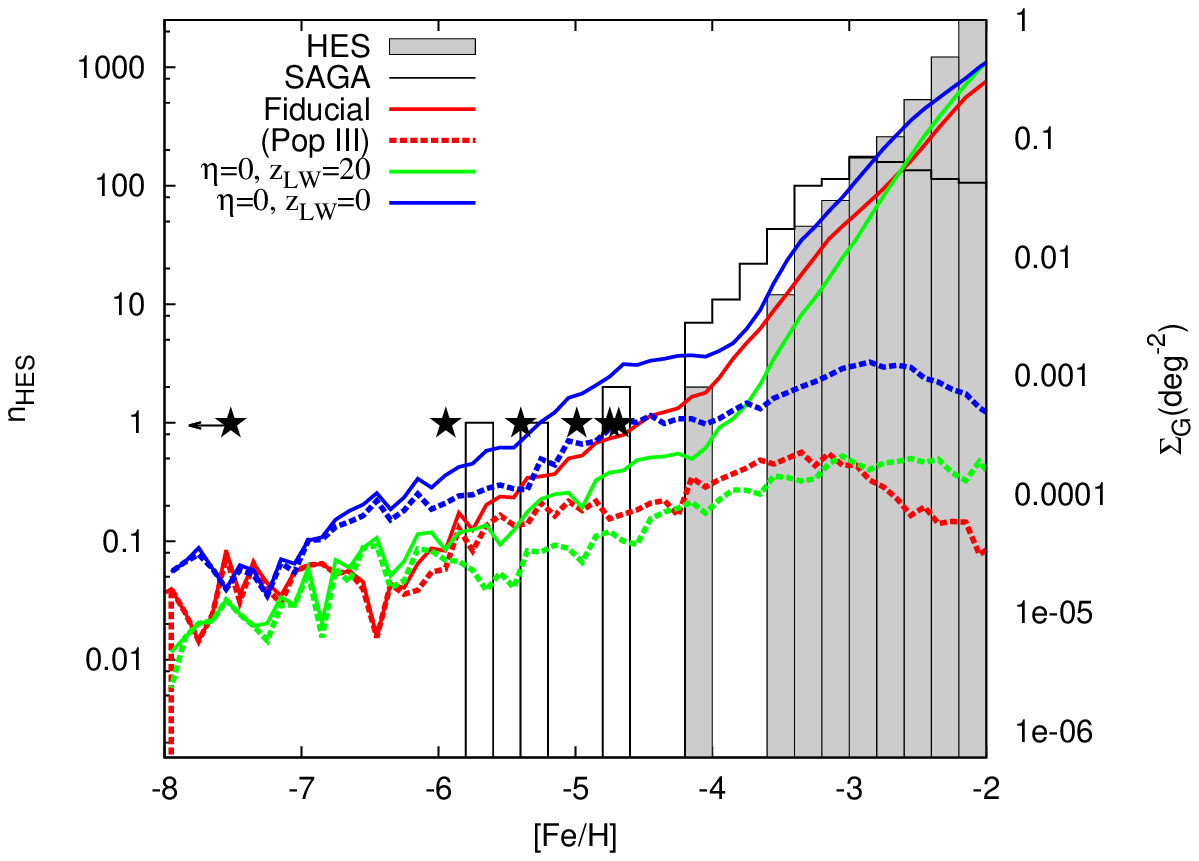}
\includegraphics[width=\columnwidth]{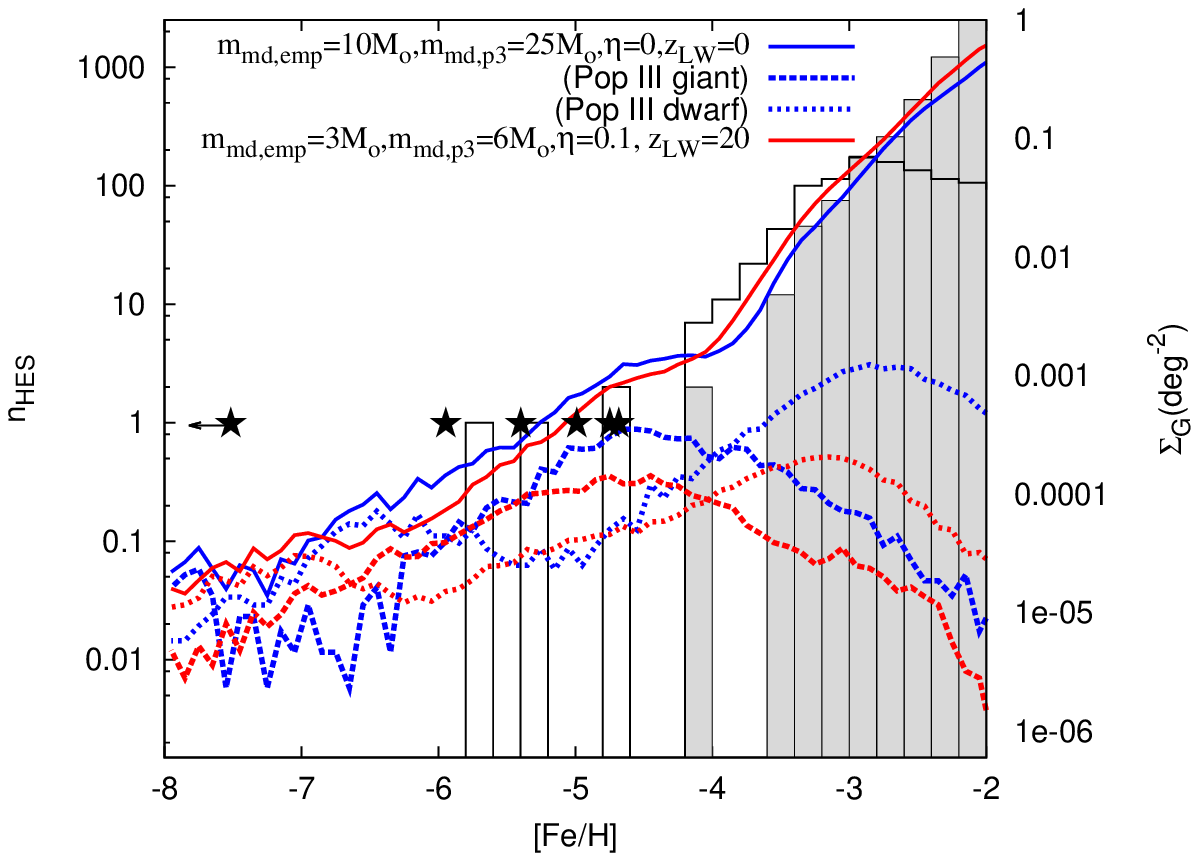}
\caption{
The predicted metallicity distribution functions (MDFs) compared with the observations (hisograms). 
   Star symbols denote the abundances with 3D corrections for stars with $\feoh < -4.5$. 
Left and right axes represent the observed number and surface number density of stars, respectively. 
{\it Top panel}: 
The predicted MDFs of all stars (solid line) and the polluted Pop~III stars (dashed line) for the fiducial model (red) and models without SN-driven outflow (green) and without SN-driven outflow and Lyman-Werner feedback (blue). 
{\it Bottom panel}: 
The same as the top panel but for the intermediate-mass IMF model with the feedback and the de Vaucleur profile (red) and the high-mass IMF model without the feedback effects (blue). 
   Dashes and dotted lines denote the MDFs of Pop~III giants and dwarfs, respectively.  
See text for details. 
}\label{MDF}
\end{figure}

Figure~\ref{MDF} shows the predicted MDFs for the polluted Pop~III and EMP stars, and the observed samples taken from the SAGA database \citep{Suda08, Suda11} and from the Hamburg/ESO (HES) survey \citep{Schorck08}.  
The SAGA data has more accurate metallicity than \citet{Schorck08} due to higher resolution, though biased toward lower metallicity of $\feoh \lesssim -3$.

In our previous study, we computed the MDF without the effects of LW feedback and SN-driven gas outflow \citep{Komiya10}. 
   The top panel shows the dependences on these feedback effects. 
The LW feedback decreases the number of stars by prohibiting star formation in small mini-halos without metals. 
   SN-driven gas outflow also decreases the number at $\feoh > -2.5$ because it reduces the mass of star forming gas. 
   For $\feoh < -2.5$, however, the number of stars rather swells as a result of pre-enrichment. 
   In addition, in the $\eta=0$ model, a majority of dwarf EMP stars becomes $\feoh > -3$ due to efficient surface pollution.

We compare the absolute number of stars predicted with the HES data considering its survey area and selection efficiency. 
The fiducial model predicts a smaller number of EMP survivors, while the model without the feedbacks gives the number in agreement with observations when we assume a homogeneous surface density of EMP halo stars. 
Additionally, if we adopt the de Vaucouleur density profile for the stellar halo, the scaling factor can be smaller by a factor of five.  
   It is to be noted, however, that the number of EMP survivors is determined by the production ratio of low- to high-mass stars since the latter dominates the chemical evolution. 
   In our models, it depends on the $\mmd$ and $\sigma $ of the IMF, the binary fraction $f_b$,  and the MRD. 
   For example, we present the result with the intermediate-mass IMF, $m_{\rm md, emp} = 3\msun$ and $m_{\rm md, p3} = 5\msun$, in the bottom panel. 
This model results in the number of survivors larger by 0.7 dex, compatible with the observation under the de Vaucouleur density profile.  
   The shape of the MDF has little dependence on the adopted IMF. 
   In all cases, stars with $\feoh < -5$ are polluted Pop~III stars except for a small contribution from the first generation stars in pre-enriched halos.

\begin{figure}
\includegraphics[width=\columnwidth]{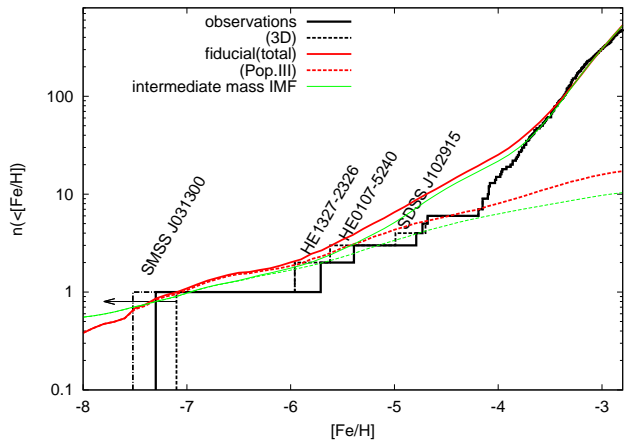}
\caption{
The cumulative MDF of all stars (solid line) and polluted Pop~III stars (dashed line) for the fiducial model (red) and the intermediate-mass IMF model (green). 
   Histograms represent the cumulative numbers of observed stars taken from the SAGA database and from \citet{Keller14} and \citet{Hansen14}. 
   Dashed line is based on the abundances with 3D corrections, and dash-dotted line is the 3D non-LTE abundance by \citet{Bessell15}. 
   The model results are scaled to match the observations at $\feoh = -3$. 
}\label{MDFcum}
\end{figure}

Figure~\ref{MDFcum} shows the cumulative MDFs. 
   The observed data are taken from the SAGA sample, and added from \citet{Keller14, Hansen14} and \citet{Bessell15} to complete the HMP/UMP sample. 
 We also show the abundances with 3D corrections when available. 
The broad distribution of polluted Pop~III survivors agrees with the distribution of the observed stars at the lowest metallicity range. 
The metallicity range of the HMP stars including \keller\ is covered by the polluted Pop~III survivors. 
We estimate the iron abundance of \keller\ to be around $\feoh \simeq -7.58$ from the observed calcium abundance ($\abra{Ca}{H} = -7.26$) and the typical enhancement of calcium for EMP stars ($\abra{Ca}{Fe} \simeq +0.32$).  
The iron abundance is unlikely to be $\feoh < -8.0$ in our scenario because $98\%$ of EMP stars show $\abra{Ca}{Fe} < +0.74$. 
Note that the expected number of observed stars is $\sim 1$ at $\feoh < -7.0$, $\sim 0.6$ at $\feoh < -7.6$, and $\sim 0.4$ at $\feoh < -8.0$. 
We see a slight overabundance for the model results around $\feoh \simeq -5$ to $-4$. 
   This is an artifact resulting from the assumption of homogeneous mixing of SN ejecta in such massive halos as $M_{\rm h} > 10^7 \msun$, and can be alleviated by considering possible inhomogeneous mixing.

We also check the dependencies on other model parameters. 
There are minor corrections to the fiducial model by changing $\dot{m_1} / \dot{m_2}$. 
Binary parameters such as $f_b$ and $n(q)$ affect the number of Pop~III and EMP survivors but have little effect on the metallicity distribution.  
On the other hand, the MDF at the lowest metallicity is affected by the SFE; 
   we find a rapid decline at $\feoh \lesssim -6$ for $\epsilon_{\star}=10^{-10}{\rm yr}^{-1}$, because of the higher metallicity that the ISM reached before the merger events.

\section{Conclusions and Discussion}\label{Sconclusion}

We have explored the change in the surface iron abundance of low-mass survivors of Pop~III stars by  the accretion of ISM using the chemical evolution model within the framework of hierarchical structure formation. 

Our model can reproduce the MDF observed for Galactic halo stars with $\feoh < -2.5$.  
   The MDF of polluted Pop~III giants peaks around $\feoh \simeq -5$ and extends to cover the whole metallicity range of HMP stars including \keller. 
   A wide spread of metallicity of polluted Pop~III stars arises from the intervals from the birth to the merger of host halos and the mass of primary stars. 
   The results little depend on the adopted IMF, and the binary parameters, but are dependent on the SFE.

Surface pollution is most effective in the mini-halos of small mass because the Bondi-Hoyle accretion rate is strongly dependent on the relative velocity of stars to the ambient gas \citep[e.g.,][]{Suda04, Komiya09}.  
   This explains the difference from \citet{Frebel09}, who considered only the surface pollution during passage through the MW disk.   
   \citet{Johnson11} argued the suppression of ISM accretion by alleged solar-like winds from Pop~III survivors. 
   However, little is known about the mass loss from Pop~III stars with a complete lack of metals and also in the possible absence of magnetic fields. 
   In addition, their argument on the dynamics of Pop~III survivors in the mini-halos may not be relevant for those belonging to binaries with more massive stars. 
   We note that \citet{Hattori14} report possible observational evidence for the surface metal pollution, finding that the median metallicity of halo G-type dwarfs may be systematically higher than that of K-type dwarfs with the same kinematics. 

The present results indicate that such small abundances of iron group elements as observed from HMP stars are readily explained in terms of the ISM accretion.  
  It is true that HMP/UMP stars show various peculiar abundance patterns other than the large enhancement of carbon, i.e., the enhancement of N, O, and light elements (Na, Mg, and Al), and also of Sr. 
   These elements can be synthesized and brought to the surface in low- and intermediate-mass stars during the AGB phase \citep{Nishimura09,Yamada15}, and transferred onto their low-mass secondary companions \citep{Suda04, Komiya07}.  
   We will discuss the relevance of the nucleosynthesis and material mixing in AGB stars to the observed abundance patterns in a subsequent paper \citep{Suda15}.

  As an alternative scenario for the origin of HMP stars, 
it has been argued that they are the stars formed from the matter polluted with the ejecta from ``faint SNe'' with large C/Fe ratios \citep[e.g.][]{Umeda03, Iwamoto05}. 
  \citet{Keller14, Ishigali14, Marassi14}, and \citet{Bessell15} argue that the abundance pattern of \keller\ is consistent with the ejecta of a peculiar SN with small (or no) iron ejection. 
   However, this scenario demands manipulation of reducing the ejection of iron by many orders of magnitude ($\sim 10^4 $-th or less). 
   It must be noted that large C/Fe ratio results solely from the reduction of iron ejecta in ``faint SNe'' since the amount of carbon produced hardly differ from ordinary SNe.  
   There should be an account for producing such large carbon abundances and variations as $\abra{C}{H} = -1.3 \mhyph -2.4$ observed for HMP stars with almost similar amount of carbon ejected \citep{Suda04, Komiya07}.

In recent simulations of first star formation, low-mass stars can be formed at the same sites as massive stars \citep[e.g.][]{Clark08, Greif11, Smith11, Dopcke13, Stacy14}. 
   This opens a new path for low-mass Pop~III star formation in addition to ionized primordial gas in the mini-halos with $T_{\rm vir} \gtrsim 10^4$ K \citep{Uehara00}.  
   Our scenario indicates that Pop~III survivors, found in the Milky Way halo, should have suffered more or less from surface pollution by the ISM accretion. 
   In order to confirm the formation of Pop~III low-mass stars, therefore, we have to seek for un-polluted Pop~III survivors, which are expected to have been expelled from the host halos.  
   We will discuss the search for these pristine Pop~III stars in a forthcoming paper.

\vspace{15pt}

This work is partially supported by Grant-in-Aid for Scientific Research (23224004, 25400233, 25800115).

\end{document}